\documentclass[a4paper]{article}
\usepackage{INTERSPEECH2015,amssymb,amsmath,epsfig,bm}
\usepackage{mathrsfs}
\usepackage{cite}
\usepackage{subfigure}

\title{Non linear time compression of clear and normal speech at high rates}

\makeatletter
\def\name#1{\gdef\@name{#1\\}}
\makeatother \name{ {\em Cassia Valentini-Botinhao$^1$, Mirjam Wester$^1$, Junichi Yamagishi$^{1,2}$} \\ {\em Markus Toman$^3$, Michael Pucher$^3$, Dietmar Schabus$^3$}}

\address{
 $^1$ The Centre for Speech Technology Research (CSTR), University of Edinburgh, UK \\
 $^2$ National Institute of Informatics, Japan \\
 $^3$ Telecommunications Research Center Vienna (FTW), Austria \\
{\small \tt \{cvbotinh,mwester,jyamagis\}@inf.ed.ac.uk, \{toman,pucher,schabus\}@ftw.at}}

\begin{document}
\ninept

\renewcommand{\baselinestretch}{0.93}
\maketitle

\begin{abstract}
We compare a series of time compression methods applied to normal and clear speech. First we evaluate a linear (uniform) method applied to these styles as well as to naturally-produced fast speech. We found, in line with the literature, that unprocessed fast speech was less intelligible than linearly compressed normal speech. Fast speech was also less intelligible than compressed clear speech but at the highest rate (three times faster than normal) the advantage of clear over fast speech was lost. To test whether this was due to shorter speech duration we evaluate, in our second experiments, a range of methods that compress speech and silence at different rates. We found that even when the overall duration of speech and silence is kept the same across styles, compressed normal speech is still more intelligible than compressed clear speech. Compressing silence twice as much as speech improved results further for normal speech with very little additional computational costs.
\end{abstract}

\section{Introduction}
\label{sec:intro}

\begin{figure*}[t]
\begin{minipage}[b]{1.0\linewidth}
  \centering
\includegraphics[width=16cm]{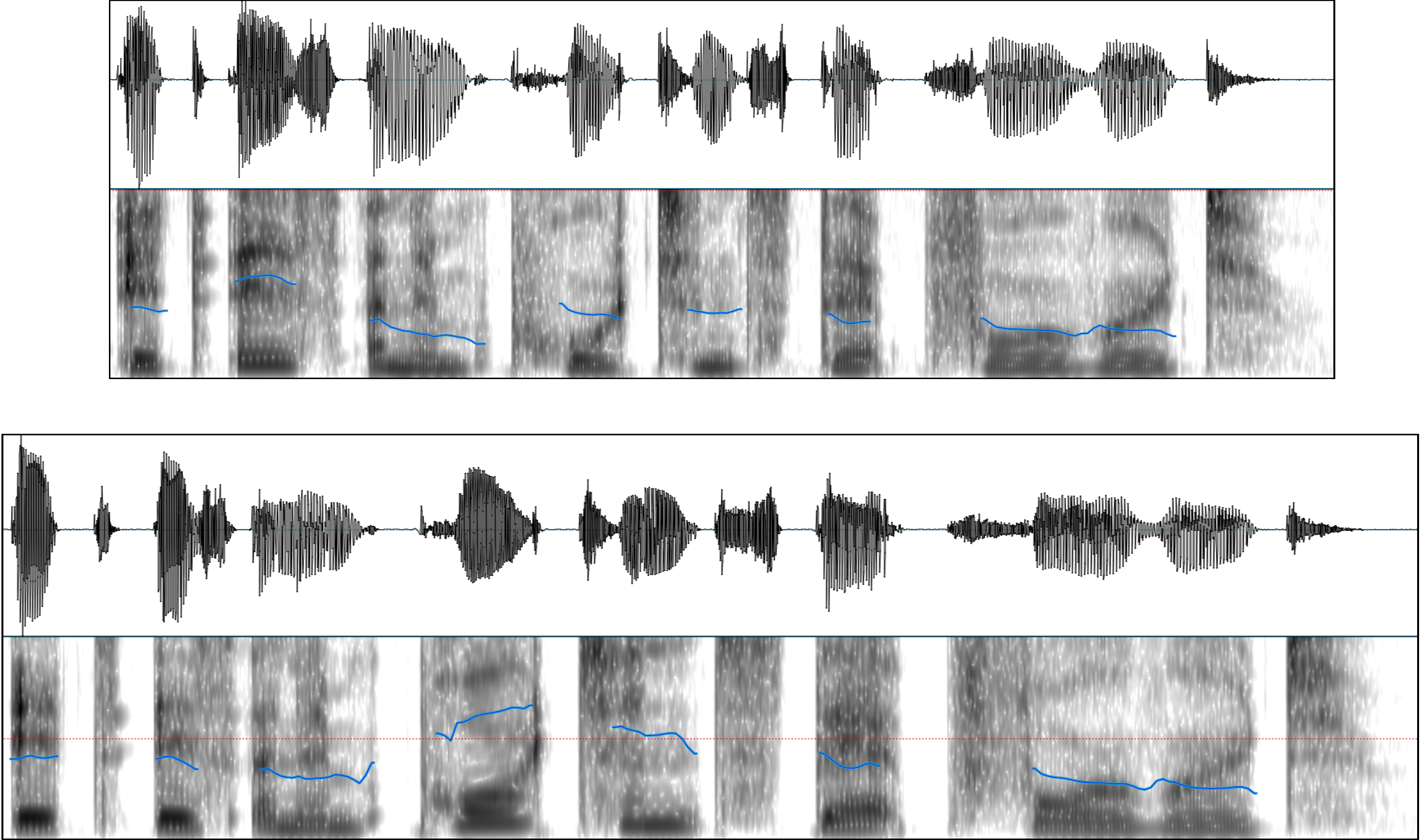}
\end{minipage}
 \caption{{\it Praat window with waveform and spectogram for the same sentence in fast (top) and clear h (bottom).}} \label{fig:spectogram}
\end{figure*}

Achieving fast speech with high levels of intelligibility is an elusive goal  \cite{Picheny89,Uchanski96,Janse03, Janse04,Krause02}. Nevertheless, there are compelling reasons why one would want to achieve it. For instance, speeding through recordings of long meetings to quickly obtain relevant content  \cite{Arons97,Tucker06}, or as a speech output interface for blind text-to-speech (TTS) users \cite{Pucher10,IS14} .  Furthermore, understanding what makes speech more intelligible can also bring improvements to hearing aids \cite{Picheny85,Picheny89,Uchanski96}.

During the production of fast speech there is an increasing  amount of overlap of articulatory gestures which results in a decrease in intelligibility, as the articulatory targets, important for clear pronunciation, are no longer reached. When producing fast speech, vowels are compressed more than consonants \cite{Gay78} and both word-level \cite{Janse03} and sentence-level \cite{Port69} stressed syllables are compressed to a lesser degree than unstressed ones. Yet another important aspect of fast speech is the significant reduction in pauses. It is claimed that reducing pauses is possibly the strongest acoustic change when speaking faster \cite{Goldman68}, most probably due to the limitations of how much speakers can speed up their articulation rate \cite{Greisbach92}. 

Janse and colleagues \cite{Janse03,Janse04} have shown that fast speech (approximately $1.56$ times faster than normal speech) is harder to process, in terms of reaction times, and is less preferred than linearly compressed speech. Following the literature, linearity here refers to the fact that the compression rate is the same across the sentence, i.e. vowels, consonants, silence and speech are compressed at the same rate. Furthermore in \cite{Janse03}, Janse and colleagues found that linearly compressed speech is more intelligible and preferred over a non linearly compressed version in which fast speech prosodic patterns were mimicked at a high speaking rate ($2.85$ times). \cite{Moers10} reported that linearly speeding up normal rate sentences to a fast rate led to more intelligible sentences. Speeding both the natural and fast speech further (ultra-fast) resulted in comparable levels of intelligibility with the fast speech being rated as more natural. Similarly, the results in \cite{IS14} show that linear compression of natural plain speech leads to higher intelligibility rates than natural fast speech.

Janse claims in \cite{Janse04} that possibly the only non linear aspect of fast speech duration changes that can improve intelligibility at high speaking rates is the removal of pauses but only when rates are relatively high, i.e., non linear compression (compressing pauses more than speech) at high speaking rates (faster than fast speech).  Results obtained using the MACH1 algorithm \cite{Covell98} confirm this.  The MACH1 method is  based on the acoustics of fast speech with the addition of compressed pauses. At ultra fast speaking rates ($2.5$ and $4.1$) MACH1 improves comprehension and is preferable to linearly compressed speech, however, no advantage was found at a fast speech speaking rate ($1.4$) \cite{He01}. 
\cite{Demol05} proposed a non uniform time scaling of speech based on the waveform similarity overlap and add (WSOLA) time compression method \cite{Verhelst93} where pauses, vowels, phone transitions and consonants are compressed differently (order is from more to less). Authors reported a preference for the non linear compression method over the linear compression method and natural fast speech.

All the above studies show that time compression methods using fast speech do not tend to lead to higher intelligibility scores than when speech at normal speaking rates is compressed. At high rates non linear compression of normal speech shows some promise. 

A possible alternative to using fast speech is approaching the problem from the other direction, i.e., by using clear speech as the basis for compression\cite{Uchanski08,Hazan11}. Clear speech is a speaking style adopted
when speaking in difficult communication situations. Clear speech is significantly more intelligible than conversational speech (particularly for individuals with some sort of hearing impairment) but at the expense of longer utterance duration \cite{Picheny85}.  Studies \cite{Picheny89,Uchanski96} testing linear and non linear time compression of clear speech found that for both compression methods clear speech reproduced at a conversational speaking rate was not more intelligible than conversational speech. However, \cite{Uchanski96} did find that non uniform time compression was less deleterious to the intelligibility of clear speech than a linear method but both types of compressed clear speech were still no more intelligible than unprocessed conversational speech; differences were larger for hearing impaired listeners. Krause and Braida \cite{Krause02} thought that this might be a problem with the compression method and that the intelligibility advantage of clear speech is not only due to its longer duration. They \cite{Krause02} investigated whether clear speech produced at high speaking rates could still bring intelligibility benefits over conversational speech and found that it does. \cite{Kouts12} also found that clear speech which was linearly compressed to match casual speech speeds was more intelligible than unmodified casual speech.

In this paper, we are interested in testing whether the clear speech advantage still holds for particularly high speaking rates and for normal hearing individuals, aiming to reproduce such results for the generation of synthetic speech to be used by blind individuals.  The questions this paper sets out to answer are: can compressed clear speech be more intelligible than compressed normal speech, and can we improve results by applying a simpler non linear technique that compresses speech and silence regions at different rates?

The remainder of this paper is as follows: Section \ref{sec:database} presents the database used in the experiments, Section \ref{sec:linear} shows results of linear compression applied to a range of speaking styles at a range of speaking rates, Section \ref{sec:nonlinear} describes the evaluation of non linear time compression methods applied to clear and normal speech at the highest rate. This is followed by a discussion and conclusions.

\section{Database}
\label{sec:database}

We recorded a Scottish female voice talent reading prompts presented sentence by sentence. The same $400$ sentences were read in four styles: normal, fast and two types of clear. Each style was elicited by different instructions. For the normal style we asked the voice talent to speak as she would normally do. For the fast style she was asked to read the sentences out load as fast as she could while still maintaining intelligibility. To create the two types of clear speech she was instructed to speak as if talking to someone with an hearing impairment (clear h) and to a computer (clear c). To illustrate how the database looks like Fig.\,\ref{fig:spectogram} shows the waveform and spectogram in Praat for the same sentence spoken with the fast and clear h style~\footnote{Speech samples used in the evaluation can be found at: https://wiki.inf.ed.ac.uk/CSTR/ClearSpeech}.

These instructions led to speech with a wide range of timing properties. Table \ref{tab:data} presents, for each speaking style, the syllables per second (SPS), words per minute (WPM), speech duration ($\Delta_{sp}$) and silence duration ($\Delta_{sil}$). All values are calculated per sentence and averaged across sentences. The values of SPS and WPM were based on a manual annotation of part of the data and consider the whole utterance including pauses, while the other values were calculated automatically by using an energy based speech detection method \cite{Demol05}.

\begin{table} [t]
\centerline{
    \begin{tabular}{ c c c c c }
     \hline \hline
        style & SPS & WPM & $\Delta_{sp}$ (secs.) & $\Delta_{sil}$ (secs.) \\ \hline
        normal & 4.65 & 215.0  & 1.63 & 0.14 \\
        	fast     & 8.10  &  371.8 & 0.97 & 0.02 \\
	    clear h & 3.32  &  151.6 & 2.13 & 0.43  \\
	    clear c & 2.05  &  93.0 & 2.88 & 1.16 \\
        \hline\hline 
    \end{tabular}}	
    \vskip0pt
\caption{{\it Syllables per second (SPS), words per minute (WPM), speech duration ($\Delta_{sp}$) and silence duration ($\Delta_{sil}$) calculated per sentence and averaged across sentences for each speaking style.}}
\label{tab:data}
\vskip-10pt
\end{table}

Table\ref{tab:data} shows that the slowest style is the clear c with $2.05$ SPS, followed by clear h with $3.32$, normal with $4.65$ and fast with $8.1$.  The duration of speech and silence inform how much of these differences are due to speech regions being longer or due to longer silence regions. For this analysis, we take normal as the reference. Silence regions not only include pauses but also phone regions such as the burst that takes place during stops.
The rate of increase of speech and silence duration is of $1.3$ and $3.1$ for clear h and $1.76$ and $8.29$ for clear c. The rate of compression of fast speech and silence duration is of $1.68$ and $7$. These values show that the overall durational differences seen in the clear and fast speech style recorded here were in fact due to silence regions being stretched and compressed, respectively.

\section{Linear time compression}
\label{sec:linear}

Fast speech has been reported to be less intelligible than linearly compressed normal speech. Here we explore whether the same holds true when compressing other speaking styles. We evaluate the intelligibility of speech compressed using the waveform similarity overlap and add (WSOLA) time compression method \cite{Verhelst93} to illustrate a linear, also referred to as uniform, compression as was evaluated in \cite{Demol05,Kouts12,IS14}. 

In this section, we compress the four different speech styles described in the previous section at four different rates. The rates chosen for this experiment are: fast (the rate that brings each style's duration to match the duration of the fast speech) as well as $2$, $2.5$ and $3$ times faster than normal speech. Intelligibility results for linearly compressed normal and fast speech for the same speaker have previously been reported in \cite{IS14}.

\subsection{Listening experiment}

Twenty native English speakers with no self reported hearing impairment participated in this experiment. Each individual listened to eight different sentences for each of the $16$ conditions tested and had to type the words they understood sentence by sentence. Prior to the test they undertook a small training session containing one example of each condition. 

\subsection{Results}

\begin{figure}[t]
\begin{minipage}[b]{1.0\linewidth}
  \centering
\includegraphics[width=8cm]{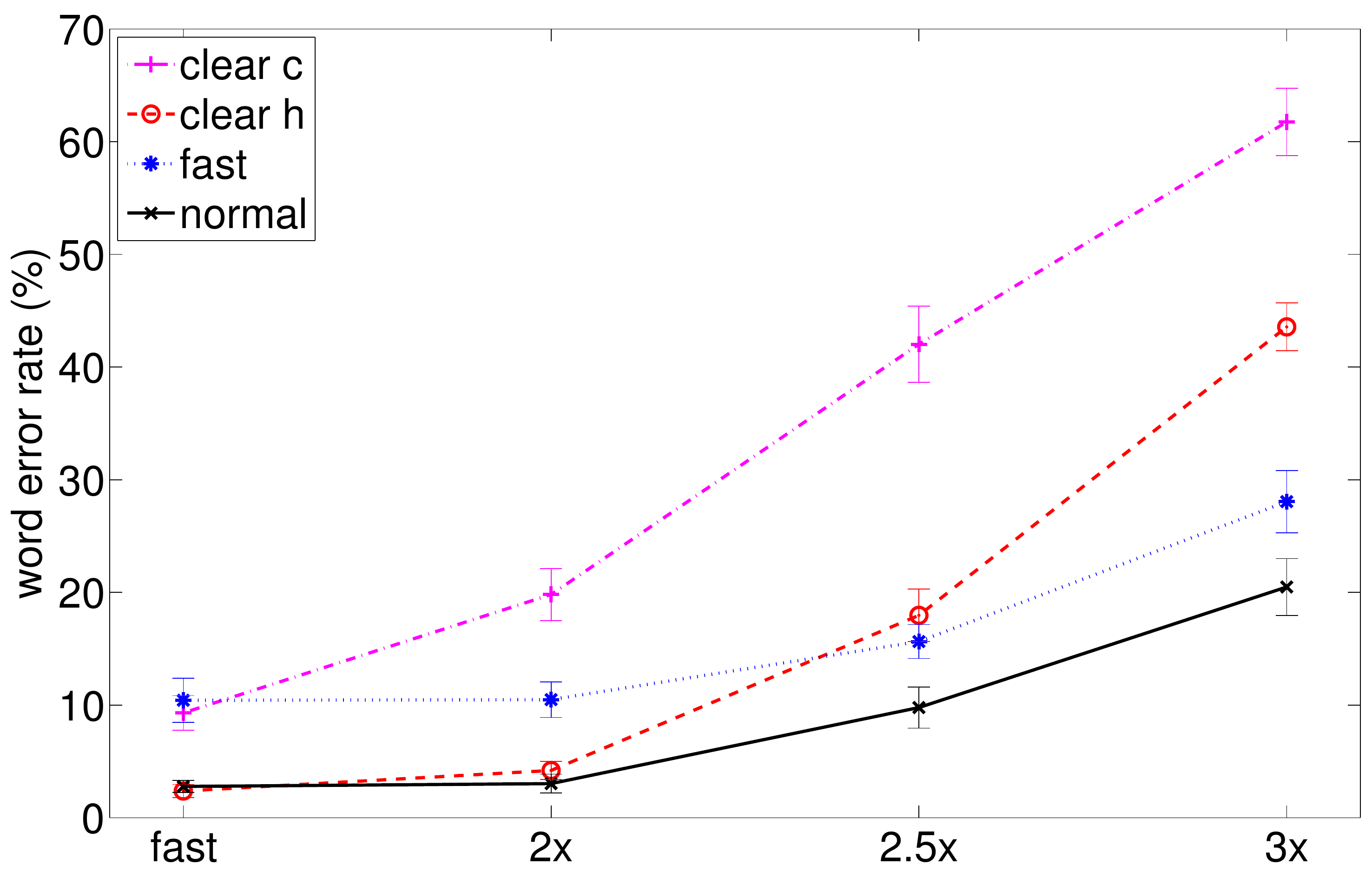}
\end{minipage}
 \caption{{\it Word error of linearly compressed speech for different speaking styles, rate is relative to normal speech, the fast rate is on average $1.55$x faster than normal speech.}} \label{fig:exp1}
\end{figure}

The results are calculated as the percentage of word errors averaged across a listener, taking into consideration misspellings and word contractions. Word errors are counted as words that did not appear in the transcription, irrespective of their placement as done in \cite{Hurricane}. Fig.\ref{fig:exp1} shows the word errors for each speaking style at each compression rate. Error bars refer to the standard deviation of the error calculated across listeners.

The most intelligible style for all rates was the normal style, leading to only $20.5$\% word errors for the highest speaking rate. The least intelligible style was the computer directed clear speech, which already produced a similar error at the 2x rate and more than $60$\% word errors at the 3x rate. For moderate speaking rates (fast and 2x) the most intelligible style, after normal, was the hearing impaired directed clear speech (clear h), with less than $5$\% errors for the 2x rate, while for higher rates fast speech becomes more intelligible, with $28$\% for the 3x rate where clear h obtained $43.6$\%. At the fast speech rate, linearly compressed normal speech was more intelligible than uncompressed fast speech, which supports the findings in \cite{Janse03,Janse04}. We found that at this rate linearly compressed clear h was also more intelligible than unprocessed fast speech. 

One particularly interesting finding was the fact that, at higher rates, compressed fast speech was more intelligible than compressed clear speech even though fast speech at its own rate was found to be less intelligible. In our view, there are two striking differences between the fast and clear h data: the duration of the silence and speech regions and how well articulatory targets were met. We expect that fast speech is inherently less intelligible than normal and clear speech due to substitutions and deletions that take place when one speaks fast. Therefore, we expect that the fast speech advantage at higher rates is due to highly compressed silence.

\section{Non linear time compression}
\label{sec:nonlinear}

\begin{table} [b]
\centerline{
    \begin{tabular}{ l c c c c }
     \hline \hline
        & style & compression & speech & silence \\ \hline 
        N-L     & normal & linear & yes & yes \\
        N-NL1 &  normal & non linear & yes & no \\
	    N-NL3 &  normal &non linear & yes & yes \\
        C-L      & clear h & linear & yes & yes \\
        C-NL1 &  clear h & non linear & yes & no \\
	    C-NL2 &  clear h & non linear & yes & yes* \\
	    C-NL3 &  clear h & non linear & yes & yes \\
        \hline\hline 
    \end{tabular}}
    \vskip0pt
\caption{{\it Methods evaluated. `yes' and `no' refer to being compressed or not. `yes*' refers to the condition where silence duration was compressed to match that of normal speech.}}
\label{tab:methods}
\vskip-10pt
\end{table}

Clear speech is known to be more intelligible than normal speech but in our previous experiment we found no intelligibility gains when compressing clear speech to produce speech at high rates. The fact that fast speech, even though inherently less intelligible, led to better results at the highest rate, could indicate that compressed clear speech is not as intelligible because of the presence of long silences which makes the speech duration of its compressed version much shorter.

For this experiment, we focus on improving results for compressed clear h and possibly normal speech for the higher speaking rate of 3x. For this we will exploit a range of non linear compression methods, always focusing on applying different rates to speech and silence. To calculate silence regions we apply a  silence detection algorithm based on a fixed energy threshold as done in \cite{Demol05}. For all methods, we calculate the rate per frame according to the characteristics of the current frame (speech or silent) and feed this information to the WSOLA method which calculates the best match for the next frame to overlap and add.

\subsection{Non linear methods}

Table \ref{tab:methods} presents the acronyms for the conditions we evaluate. L refers to the  condition tested in the previous experiment, i.e. compress the whole utterance (speech and silence) with the same rate (the linear or uniform method). 

We were interested in exploring whether clear speech was found to be less intelligible at higher rates because the duration of speech was shorter due to the presence of longer silence regions. To test this hypothesis we create condition NL1 where we compress speech to the same rate as in L so that speech duration is the same across styles and silence remains uncompressed. The final utterance duration will therefore be longer for NL1 than for L.

The resulting durations across speaking styles will vary as the duration of silence is different which makes the comparison of NL1 across styles unfair. The fair comparison is condition NL2 applied to clear speech only, where silence is also compressed so that the overall utterance duration is the same across N-NL1 and C-NL2. 

Finally to test the theory that at higher rates pauses harm more than aid, we compress silence Y times more than speech for both styles (N-NL3 and C-NL3). Y was chosen to be two.

\subsection{Listening experiment}

As in the previous experiment, twenty native English speakers with no self reported hearing impairment transcribed eight sentences for each condition after hearing each sentence only once. One participant was removed from the results as his/her word errors were found to be excessively high compared to the others.

\begin{figure}[t]
\begin{minipage}[b]{1.0\linewidth}
  \centering
  \subfigure{\includegraphics[width=8cm]{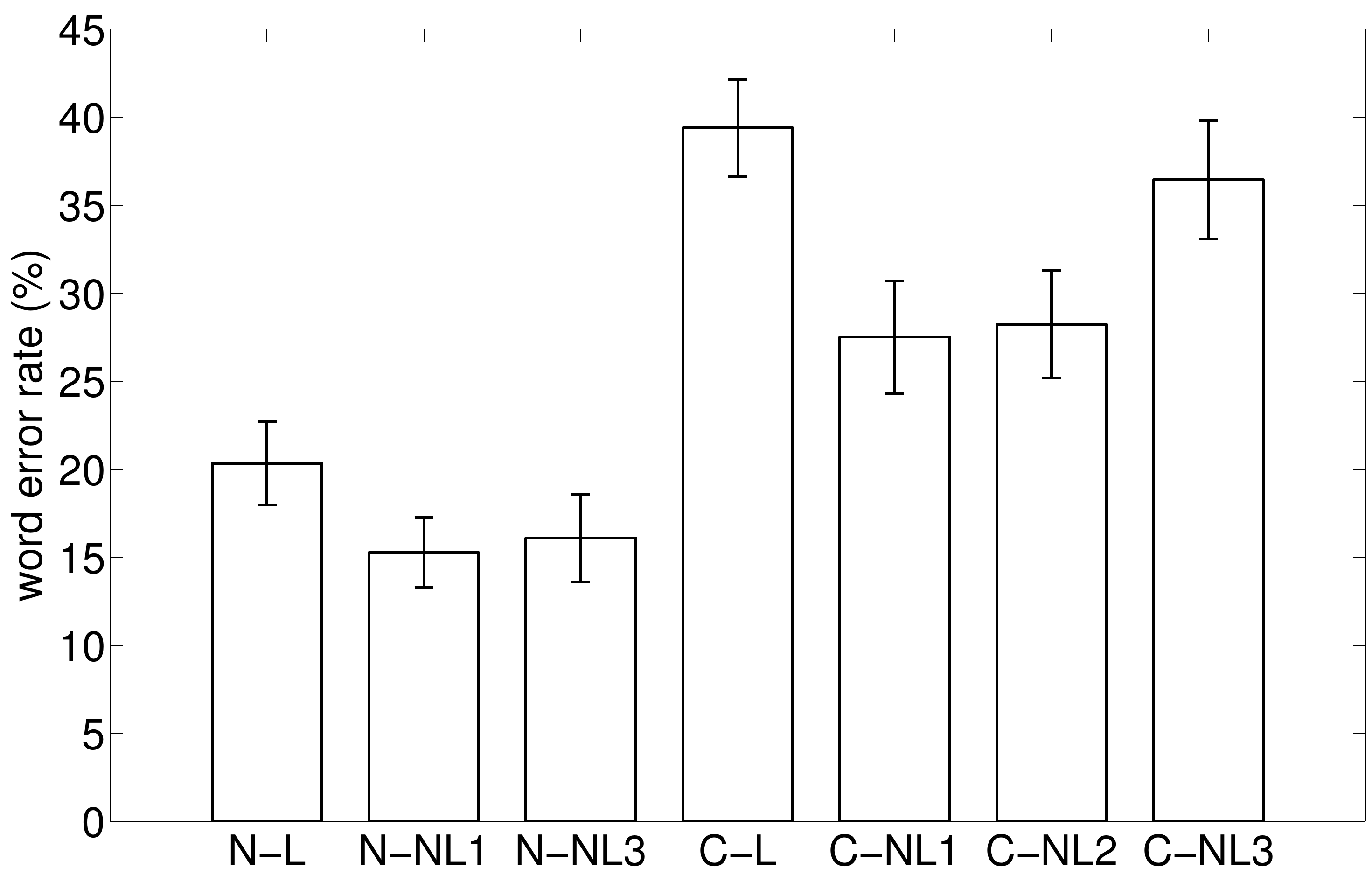}}  \label{fig:exp2_r3}
\end{minipage} \vskip-10pt
 \caption{{\it Word errors at 3x rate for normal (N-) and clear h (C-) speech compressed using a range of methods.}} \label{fig:exp2_r2}
\end{figure}

\subsection{Results}

The results are calculated as the percentage of word errors averaged across a listener, in the same way as in the previous experiment. Fig.\ref{fig:exp2_r2} presents the word error in percentage for the 3x rate.

Similar to what was found in the previous experiment the word errors for linearly compressed clear speech (C-L) $39.4$\% is around twice as high as results with linearly compressed normal speech (N-L) $20.3$\%. This relation remains the same when the final duration of speech and silence is the same across styles: $15.3$\% of N-NL1 and $28.2$\% of C-NL2.

Not compressing pauses improved results for both styles significantly as we see NL1 scores are lower than L, but at the expense of a longer utterance duration. Compressing silence twice as much as speech also improves results as we see that for both styles NL3 results are better than L, even though the utterance duration is the same. The difference was found to be significant only for the normal style. All types of non linear methods applied to clear speech resulted in intelligibility scores closer to normal speech but not equal to or better than.

\section{Discussion}
\label{sec:discussions}

We were interested in finding whether linearly compressed clear speech was found to be less intelligible at higher rates because the duration of speech was shorter due to the presence of longer silence regions. This was however not the case as when we set the duration of speech and silence to be the same for compressed normal speech and clear speech, we found that compressed normal speech was more intelligible. One possible reason is that clear speech had to be compressed considerably more than normal speech which could have caused more artefacts due to larger phase differences at frame boundaries brought upon by the compression. The WSOLA implementation used here \cite{Rabiner10} suggests no more than 4.0x the compression rate and for many sentences clear speech was compressed more than this. Future work will involve reducing such artefacts.

The non linear compression method improved results for both styles. Unfortunately, this was not enough to make compressed clear speech more intelligible than normal speech.  This is an interesting result as the overhead of applying energy detection is quite small and requires no further delay; it can be done online as opposed to methods where all silence regions are first completely removed. We would like to quantify what further intelligibility gains more complicated non linear methods inspired by fast speech acoustics \cite{Covell98,Demol05} might bring and particularly whether most of the gain they obtain are due to heavy silence compression.

Another point of discussion is why the type of clear speech used in this experiment (read clear speech) was less intelligible after time compression; this may be because it was not recorded in a communicative spontaneous task, where acoustic changes are known to be less extreme \cite{Hazan10}.

\section{Conclusions}
\label{sec:conclusions}

In this paper we exploit the use of clear speech to increase the intelligibility of speech reproduced at extremely high speaking rates. We first evaluate linearly compressed speech of four styles produced by the same speaker: normal, fast and two types of clear speech - computer directed and hearing-impaired directed speech (clear h). We found that unprocessed fast speech was less intelligible than linearly compressed clear h and normal but at the highest speaking rates clear h was worse than fast. As possibly the only advantage fast speech has over clear in terms of intelligibility is the shorter silence regions, we exploit in our second experiments a range of time compression methods that compress speech and silence differently. We found that even when the duration of speech and silence is kept the same across styles, compressed normal speech is still more intelligible. Compressing silence twice as much as speech improved results for both styles at the expense of very little overhead. 

\vspace{5mm}
\section{Acknowledgements}
This work was partially supported by the BMWF - Sparkling Science project {\it Sprachsynthese von Auditiven Lehrb\"uchern f\"ur Blinde Sch\"ulerInnen} (SALB) and  the EPSRC Programme Grant EP/I031022/1 (Natural Speech Technology).

\vfill\pagebreak

\pagebreak
\linespread{0.95}
\selectfont

  \newpage
  \eightpt
  \bibliographystyle{IEEEtran}

\bibliography{refs}

\end{document}